\documentclass[aps,prd,preprint,showpacs]{revtex4}
\usepackage{graphicx}
\usepackage{subfigure}
\usepackage{hyperref}
\usepackage{amsmath}
\begin{document}
\draft
\title{Associated production of different flavor heavy quarks through $W'$ bosons at the LHC }

\author{I. T. Cakir}
\email{ilkay.turkcakir@gmail.com} \affiliation{Ankara University,
Department of Physics, 06100, Ankara, Turkey}
\author{A. Senol}\email{asenol@kastamonu.edu.tr}
 \affiliation{Kastamonu University, Department of
Physics, 37100, Kastamonu, Turkey} \affiliation{Abant Izzet Baysal
University, Department of Physics, 14280, Bolu, Turkey}
\author{A. T. Tasci}\email{atasci@kastamonu.edu.tr}
 \affiliation{Kastamonu University, Department of
Physics, 37100, Kastamonu, Turkey}
\begin{abstract}
Associated production of different flavor heavy quarks can provide a
probe for $W'$ bosons at the LHC. We study  $pp\to t'\bar{b'}X$
process with the subsequent decays $t'\to W^+b$ and $\bar{b'}\to
W^-W^+\bar b$, and compare the results with the $t\bar{b}$ final
state. The effects of the $W'$ boson to the different flavor pair
production cross section are shown to be important in some parameter
region for the heavy quark masses of 600 and 700 GeV. We present
accessible mass limits for the $W'$ boson at the LHC with $\sqrt
s$=14 TeV.
\end{abstract}
 \maketitle
\section{introduction}
Discovery of new particles plays an important role for a clear
evidence of the new physics at the Large Hadron Collider (LHC). The
existence of additional charged massive bosons is predicted by
various extensions of the Standard Model (SM). Such models
candidates are the left-right symmetric \cite{Mohapatra:1974hk,
Mohapatra:1974gc, Senjanovic:1975rk}, extra dimensional
\cite{Agashe:2003zs, Appelquist:2000nn}, little Higgs
\cite{ArkaniHamed:2002qy, Kaplan:2003uc, Han:2003wu} and models with
extended gauge symmetry \cite{Malkawi:1996fs, Muller:1996dj,
Pisano:1991ee,Frampton:1992wt,Hoang:1995vq}.

A reference model given in Ref. \cite{Altarelli:1989ff} can be used
to confirm the results from experiments for searching new heavy
bosons. The recent experimental results have reported the exclusion
limit for the mass of the extra charged gauge boson ($W'$) through
leptonic decay channel below 2.5 TeV with an integrated luminosity
of 4.7 fb$^{-1}$ from ATLAS \cite{:2012dm} and 5.0 fb$^{-1}$ from
CMS \cite{Chatrchyan:2012meb} Collaborations at the center of mass
energy of 7 TeV.

The recent searches by the ATLAS Collaboration in the lepton+jet
final state, under the assumption of $BR(t'\to Wb)=1$ based on $4.7$
fb$^{-1}$ of data at $\sqrt{s}=$ 7 TeV excluded the existence of a
fourth generation up type quark $t'$ with a mass lower than 656 GeV
\cite{ATLAS:2012qe}. A search for the same sign dilepton signature
by the ATLAS Collaboration in 4.7 fb $^{-1}$ of data sets lower
limit on $m_{b'}>670$ GeV \cite{key2}. The CMS Collaboration using 5
fb $^{-1}$ of data in the same final state set lower limit of
$m_{t'}>570$ GeV \cite{Chatrchyan:2012vu}. A search for fourth
generation down type heavy quarks in the same sign dilepton or three
lepton events by the CMS Collaboration sets a lower mass limit at
611 GeV with 4.9 fb$^{-1}$ of data \cite{Chatrchyan:2012yea}.

There are also constraints on the fourth generation quarks from
Higgs searches at the Tevatron \cite{TEVNPH:2012ab} and the LHC
\cite{:2012gk,:2012gu}. The fourth generation fermions affect the
phenomenology of the Higgs, leading to the changes in decays
$\Gamma(H\to VV)$. The constraints of the fourth generation model
can be relaxed when there is an extended Higgs sector, such as two
Higgs doublet model \cite{Branco:2011iw}, with an extra $SU(2)$
symmetry.

In this study, we calculate cross section of associated production
of different flavor heavy quarks ($t'$, $b'$) with the effects of
$W'$ boson at the LHC with $\sqrt s$=14 TeV. We also calculate cross
section of relevant background processes. We study accessible mass
limits for the $W'$ boson for different heavy quark masses of 600
and 700 GeV at the LHC. The result of this study can also be
interpreted generically in the framework of heavy quark models in
which $BR(Q\to W q)$ becomes smaller than unity.
\section{ The Model}
The general interaction Lagrangian for $W'$ and $W$ bosons with the
SM quarks and fourth generation quarks ($t'$ and $b'$) is given by
\begin{eqnarray}
{\cal L} = -\frac{g}{\sqrt{2}}~\sum_{\substack{i= u,c,t,t' \\ j
=d,s,b,b'\\i\neq j}} [ V'_{ij}\bar{q_i} \gamma^\mu (f_LP_L
+f_RP_R)q_j~W^{\prime +}_\mu + V_{ij}\bar{q_i} \gamma^\mu P_L q_j
~W^{+}_\mu] + {\rm h.c.}~~
\end{eqnarray}
where $g=e/\sin\theta_W$ is the electroweak coupling constant,
$P_{L,R}$ are the usual chirality projection operators and $V_{ij}$
is the element of the extended CKM mixing matrix (CKM4) and
$V'_{ij}$ includes new mixing with the $W'$ boson. $W^{\prime
+}_\mu$ and $W^{+}_\mu$ are the vector field of $W'$ and $W$ bosons,
respectively. $f_L$ and $f_R$ are the couplings of quarks with the
$W'$ boson corresponding to a new $SU(2)$ symmetry
\cite{Cao:2012ng}. The triple gauge interaction of the $W'$ and SM
gauge bosons is included in the benchmark model as given by
Ref\cite{Altarelli:1989ff}.

A fit for the fourth generation quark mixing matrix elements (CKM4)
are given as $|V_{t'd}|=0.0058(0.0056)$, $|V_{t's}|=0.0343(0.0309)$,
$|V_{t'b}|=0.1155(0.1185)$, $|V_{ub'}|=0.0140(0.0130)$,
$|V_{cb'}|=0.0339(0.0309)$, $|V_{tb'}|=0.1149(0.1179)$,
$|V_{t'b'}|=0.9926(0.9924)$ for $m_{t'}=600 (700)$ GeV
\cite{Eilam:2009hz}. In this study, we take $f_R$=0 and $f_LV'_{ij}$
to be same as CKM4 elements, and we assume the mass constraint
$|m_{b'}-m_{t'}|\simeq 55$ GeV. For a simulation framework, the
interactions of four generations quarks with $W$ and $W'$ bosons are
implemented into computer package CompHEP \cite{Boos:2004kh}.

\begin{table}[hptb!]
\caption{The cross sections ($\sigma_1$ and $\sigma_2$) of the
signal process $pp\to t'\bar{b'}X$ for $t'$ mass 600 GeV and 700 GeV
as well as $SS$ values at $\sqrt s$=14 TeV with $L=10^5$ pb$^{-1}$.
Here, $SS_{ij}=S_i/\sqrt{B_j}$, $i,j=1,2$ are the signal
significances for the corresponding signal and background events.
$B_1$ and $B_2$ denotes number of backgrounds events for
$W^-W^+W^+Z$ and $W^-W^+W^+H$, respectively. }\label{tab1}
\begin{tabular}{lcccccccccccccccccccccccccccccc}
  \hline
  $m_{W'}$ (GeV) & &  & & $\sigma_{1}\times 10^{-2}$ (pb) & &  & &
$\sigma_{2}\times 10^{-2}$ (pb) & &  & & $SS_{11}$
   & & & &  & & $SS_{12}$ & &  & & & & $SS_{21}$ & &  & & & &
$SS_{22}$   \\\hline
  2000 & &  & & $16.10$ & &  & & $14.00$ & &  & & $136.2$ & & & & & &
$417.3$ & & & & & & $122.3$  & & & & & & $374.5$  \\
  2200 & &  & & $10.40$ & &  & & $9.40$  & &  & & $88.2 $ & & & & & &
$270.2$ & & & & & & $82.0 $  & & & & & & $251.3$  \\
  2400 & &  & & $6.78$  & &  & & $6.22$  & &  & & $57.3 $ & & & & & &
$175.5$ & & & & & & $54.3 $  & & & & & & $166.4$  \\
  2600 & &  & & $4.50$  & &  & & $4.11$  & &  & & $37.9 $ & & & & & &
$116.4$ & & & & & & $35.9 $  & & & & & & $109.8$  \\
  2800 & &  & & $2.96$  & &  & & $2.72$  & &  & & $24.9 $ & & & & & &
$76.6 $ & & & & & & $23.7 $  & & & & & & $72.7 $  \\
  3000 & &  & & $2.01$  & &  & & $1.81$  & &  & & $16.9 $ & & & & & &
$52.0 $ & & & & & & $15.8 $  & & & & & & $48.5 $  \\
  3200 & &  & & $1.40$  & &  & & $1.22$  & &  & & $11.8 $ & & & & & &
$36.2 $ & & & & & & $10.7 $  & & & & & & $32.7 $  \\
  3400 & &  & & $1.01$  & &  & & $0.84$  & &  & & $8.6  $ & & & & & &
$26.2 $ & & & & & & $7.4  $  & & & & & & $22.5 $  \\
  3600 & &  & & $0.76$  & &  & & $0.59$  & &  & & $6.5  $ & & & & & &
$19.8 $ & & & & & & $5.2  $  & & & & & & $15.9 $  \\
  3800 & &  & & $0.61$  & &  & & $0.43$  & &  & & $5.1  $ & & & & & &
$15.7 $ & & & & & & $3.8  $  & & & & & & $11.6 $  \\
  4000 & &  & & $0.51$  & &  & & $0.33$  & &  & & $4.3  $ & & & & & &
$13.2 $ & & & & & & $2.9  $  & & & & & & $8.9  $  \\ \hline
\end{tabular}
  \end{table}

\section{Analysis and Results}
\begin{figure*}[htbp!]
  \includegraphics[width=10cm]{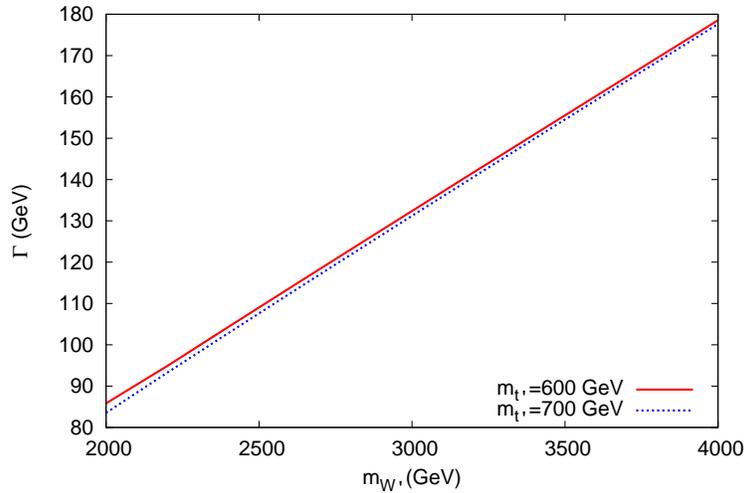}\\
 \caption{Decay widths of $W'$ boson depending on its mass for m$_{t'}=600$ GeV (solid line) and 700 GeV (dotted
 line).}\label{fig1}
\end{figure*}

The total decay widths of $W'$ boson depending on its mass
 for the mass values of $t'$ quark (600 and 700 GeV) are shown in Fig.~\ref{fig1}.
 As can be seen from this figure, the decay widths are slightly different from each
 other. The estimated branching ratios of $W'$ boson to the pair of heavy quarks are the following: $BR(W'\to
  t \bar{b})\simeq 19\% $ and $BR(W'\to
  t' \bar{b'})\simeq 16\%$. The $W'\to WZ$ mode becomes $\simeq 1 \%$ for vertex
  parametrization $\xi=M_W^2/M_{W'}^2$. These branching ratios do
  not change significantly in the considered mass range of $W'$ boson.

\begin{figure*}[htbp!]  
  \includegraphics[width=10cm]{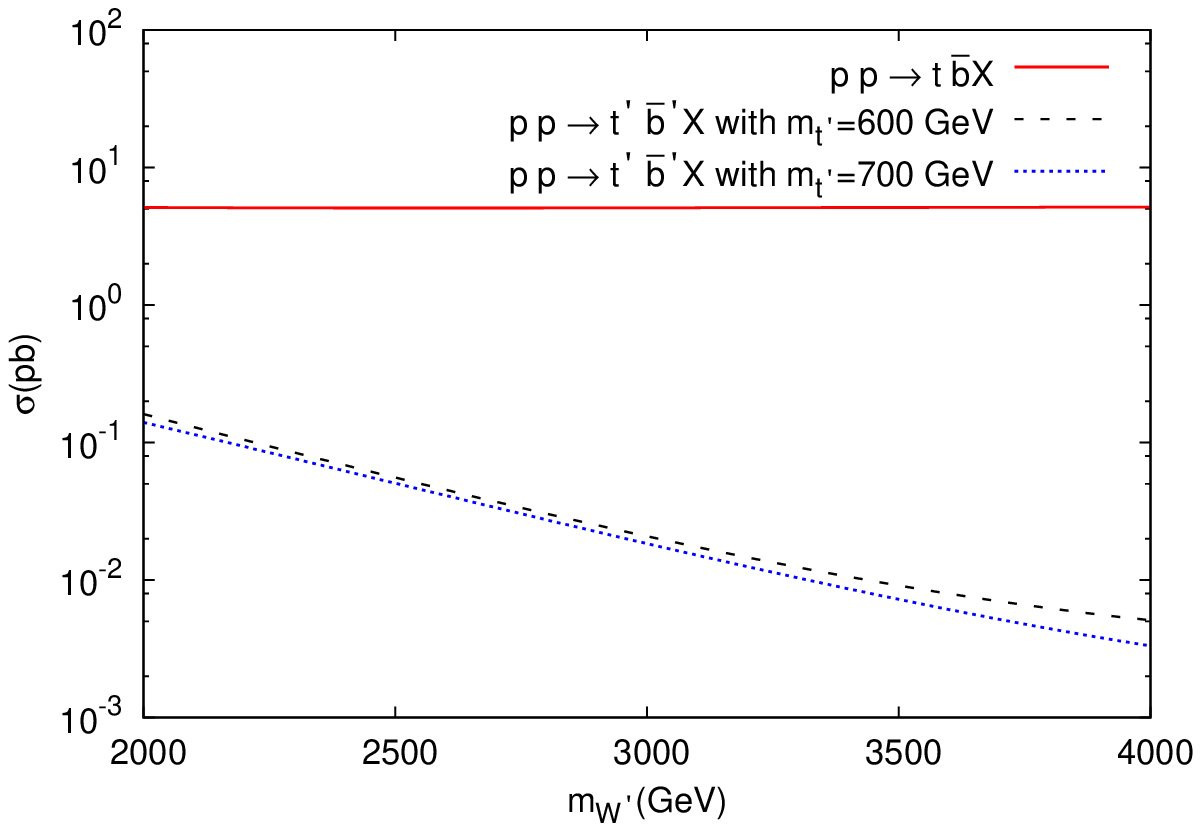}\\
 \caption{The total cross section of $pp\rightarrow t'\bar{b'}X$ for m$_{t'}=600$ GeV (dashed line) and 700 GeV (dotted
 line), and $pp\rightarrow t\bar{b}X$ (solid line) as function of $M_{W'}$ at the LHC with $\sqrt s$=14
 TeV.
 }\label{fig2}
\end{figure*}

First, we consider signal process $pp\to(W',W)\to t\bar{b}X$ at the
LHC. The cross section for the process is approximately 5.1 pb by
using parton distribution functions library CTEQ6M
\cite{Pumplin:2002vw} with $Q^2=M_{W'}^2$. Fig.~\ref{fig2} shows the
total cross sections of $pp\to(W',W)\to t\bar{b}X$ (solid line)
depending on the mass of the $W'$ bosons at the collision center of
mass energy of 14 TeV. From this figure it is seen that the cross
section sligthly changes with the mass of $W'$- boson in the range
2000-4000 GeV. For the cross section estimate, we assume the
efficiency for b-tagging to be $\varepsilon$ =0.5 and we take the
integrated luminosity $L=10^{5}$ pb$^{-1}$ at the center of mass
energy $\sqrt s=14$ TeV. In this case, the background process $pp\to
W^+ b\bar{b}X$ is calculated as $1.88\times10^2$ pb. We calculate
the statistical significance $SS=S/\sqrt B$, as 27.4 - 27.9
depending on the mass M$_{W'}$ where $S$ and $B$ denotes number of
events of signal and background, respectively. The cross section of
the process $pp\to(W,W')\to t \bar{b} X$ slightly changes with the
mass of $W'$-boson.

Second, we consider signal process $pp\to (W',W)\to t'\bar{ b'}$ at
the LHC ($\sqrt s$=14 TeV) with subsequent decays $t'\to W^+b$,
$\bar{b'}\to W^+ \bar t$ and $\bar t\to W^-\bar b$. The cross
section depending on the $W'$ mass is given in Fig.~\ref{fig2}
taking $m_{t'}=600$ GeV (dashed line) and 700 GeV (dotted line). We
take into account the primary SM background processes having the
same final state with the signal process. In this case, we have two
main background processes as $pp\to W^+W^+W^-Z$ ($B_1$) and $pp\to
W^+W^+W^-H$ ($B_2$). We calculate the cross sections of these
backgrounds as $\sigma_{B_1}=3.58\times10^{-4}$ pb and
$\sigma_{B_2}=4.44\times10^{-5}$ pb. In Table \ref{tab1}, we present
total cross sections of the signal processes, ($\sigma_1$) for
$m_{t'}$=600 GeV, ($\sigma_2$) for $m_{t'}$=700 GeV, statistical
significance for $W'$ mass range 2000-4000 GeV at the center of mass
energy $\sqrt s$=14 TeV and integrated luminosity $L=10^5$
pb$^{-1}$. When calculating the number of signal and background
events we take into account all $W$ bosons decay leptonically in the
final state, $Z$ boson and Higgs boson decay to $b\bar b$. In the
final state including $b$ quarks we take the $b$-tagging efficiency
as $\epsilon= 0.5$.

\section{conclusion}
We investigate the mass reach for $W'$ boson decaying to different
flavor heavy quarks with mass above the current mass bound at the
LHC. With an integrated luminosity of $10^5$ pb$^{-1}$ and $\sqrt
s$=14 TeV the $W'$ mass exclusion below 4000 GeV is possible for the
heavy quarks with mass 600-700 GeV. We find that $W'\to t \bar b$
and $W'\to t' \bar {b'}$ channels can be used to analyze new charged
vector boson with the exploration of the parameter space of new
charged current models.
\begin{acknowledgements}
The authors would like to thank O. Cakir for very insightful
discussions.
\end{acknowledgements}
  
\end{document}